\documentclass{ws-ijmpd}
\usepackage[super]{cite}
\usepackage{xcolor}
\usepackage[verbose,hypertexnames=false]{hyperref}
\hypersetup{colorlinks=false,allbordercolors=blue,pdfborderstyle={/S/U/W 1}}

\begin{document}

\markboth{Authors' Names}
{Neutron Star with Dark Matter Admixture: A Candidate for Bridging the Mass Gap}

%
\catchline{}{}{}{}{}
%

\title{Neutron Star with Dark Matter Admixture: A Candidate for Bridging the Mass Gap }

\author{M. Vikiaris}

\address{Department of Theoretical Physics, Aristotle University of Thessaloniki, 54124 Thessaloniki, Greece\\
mvikiari@auth.gr}

\author{V. Petousis}

\address{Institute of Experimental and Applied Physics, Czech Technical University, Prague, 110 00, Czechia\\
Vlasios.Petousis@cvut.cz}

\author{M. Veselsk\'y}

\address{Institute of Experimental and Applied Physics, Czech Technical University, Prague, 110 00, Czechia\\
Martin.Veselsky@cvut.cz}

\author{Ch.C. Moustakidis}

\address{Department of Theoretical Physics, Aristotle University of Thessaloniki, 54124 Thessaloniki, Greece\\
moustaki@auth.gr}

\maketitle

\begin{history}
\received{(Day Month Year)}
\revised{(Day Month Year)}
\accepted{(Day Month Year)}
\published{(Day Month Year)}
\end{history}

\begin{abstract}
Neutron stars, white dwarfs and black holes are the after death remnants of massive stars. However, according to the most recent observations, the neutron stars maximum mass is between $2.0-2.5 \ M_{\odot}$ while black holes of less than 5 $M_{\odot}$  has not yet been observed. The region between the most massive neutron star and the least massive black hole is called the mass-gap. If indeed its existence is confirmed by future observations, that indicates a gap in our understanding which seeks for explanation. In addition, the existence of compact objects within the mass-gap should also be supported with the help of possible new theoretical scenarios. In this study, we propose a possible explanation for the existence of compact objects within the mass-gap region. Specifically, we propose that the mass-gap region could be bridged by the existence of a hybrid compact object, composed of hadronic and self interacting and  non-annihilating fermionic dark matter, considering that the interaction between these two fluids it’s only gravitational. Fundamental questions about how these objects form and how they can be detected are also addressed.
\end{abstract}


\section{Introduction}
Astrophysical compact objects come in three varieties: neutron stars (NSs), white dwarfs (WDs) and black holes (BHs), which are the remnants of massive stars that reach the end of their lives with their cores collapsing in a supernova explosion~\cite{Shapiro-1983,Glendenning-2000,Haensel-2007}. NSs are physical nuclear laboratories with complex internal composition with main components neutrons, protons and electrons but also some exotic particles including hyperons, pions and kaons,  muons or even deconfined quark matter. BHs, which  are formed when massive stars collapse under their own gravity,  are objects with a very strong gravity characterized mainly by the three quantities that is the mass, the spin and the charge. 
Both of these entities, NSs and BHs, have uncertain mass limits. 

In a pioneering idea of Rhoades and Ruffini an   upper bound of mass of non-rotating NS was derived using a variations and conditions of standard General-Relativity equation of hydrostatic equilibrium, the Le Chattelier’s principle and the principle of causality~\cite{Rhoades-1974}. This maximum mass of the equilibrium configuration of a NS cannot be larger than $3.2 \ M_{\odot}$\cite{Rhoades-1974}.
The problem of placing an upper bound on the mass of non-rotating NS was considered also by other authors (see~\cite{Hartle-1978} and references therein) adopting similar assumptions to those
of Ref.~\cite{Rhoades-1974}. However, the majority of  nuclear matter equations of state (EoS), and which have been developed based on all possible assumptions regarding the composition of NSs, predict that their maximum mass must be  between $2.2-2.5 \ M_{\odot}$~\cite{Shapiro-1983}. Moreover, the authors in  Ref.~\cite{Rezzolla-2018},
combining the gravitational waves (GW) observations of merging systems of binary neutron stars (exploiting the recent observation of the GW event GW170817~\cite{ Abbott-2017} and drawing from basic arguments on kilonova modeling of GRB 170817A) and quasi-universal relations, set constraints on the maximum mass that can be attained by nonrotating stellar models of neutron stars. More specifically, they set limits for the maximum mass to be $2.01_{-0.04}^{+0.04} \leq M_{\rm max}/M_{\odot}\lesssim  2.16_{-0.15}^{+0.17}$. In Ref.~\cite{Shibata-2019} the authors revisited the constraints on the maximum mass of cold spherical neutron stars coming from the
observational results of GW170817 and they found that  the maximum mass of cold
spherical neutron stars can be only weakly constrained as $M_{\rm max}\lesssim 2.3 \ M_{\odot}$. In the same framework, the authors in Ref.~\cite{Khadkikar-2021} used  the GW170817 event by adopting the scenario
in which a hypermassive compact star remnant formed in a merger evolves into a supramassive compact star
that collapses into a black hole once the stability line for such stars is crossed. They deduced an upper limit on the
maximum mass of static, cold neutron stars   $2.15_{-0.17}^{+0.18} \leq M_{\rm max}/M_{\odot}\leq  2.24_{-0.44}^{+0.45}$ for the typical range of entropy
per baryon, $2\leq S/A\leq 3$ and electron fraction $Y_e=0.1$ characterizing the hot hypermassive star. 

In addition, BHs of less than $5 \ M_{\odot}$ has not yet been observed. These limits suggest a “mass gap” between the most massive NS and the least massive BH.  More specifically, in the study~\cite{Ozel-2010}, the authors, having conducted an extensive study, concluded that there is a paucity of BHs with masses $\simeq 2-5 \  M_{\odot}$. Their results confirmed and strengthen an earlier finding
of Bailyn et al.~\cite{Bailyn-1998}, who argued for a low-mass gap based
on a limited sample of BHs. Moreover, Farr et al.~\cite{Farr-2011} examine the existence of a gap between the
most massive NSs and the least massive BHs by considering the value, $M_{1\%}$, of the $1\%$ quantile
from each BH mass distribution as the lower bound of BH masses. Their analysis generates posterior
distributions for $M_{1\%}$ where the best model (the power law) fitted to the low-mass systems has a distribution of lower
bounds with $M_{1\%} > 4.3 \  M_{\odot}$ with $90\%$ confidence, while the best model (the exponential) fitted to all 20 systems
has $M_{1\%} > 4.5 \  M_{\odot}$ with $90\%$ confidence. They concluded that their sample of BH masses offers compelling evidence of a gap between the maximum NS mass and the lower limit of BH masses~\cite{Farr-2011}. On the other hand however, in Ref.~\cite{Ye-2024} the authors model the dense globular cluster NGC 1851 with a grid of similar dense star clusters utilizing the state-of-the-art Monte Carlo N-body code Cluster Monte Carlo, and they specifically studied  the formation of LMG-BHs. They demonstrated that both massive star evolution and dynamical interactions can play a role in the formation of low-mass black holes.  
 It is worth to mention here that in the analysis of the  multimessenger event GW170817–GRB170817A, in Ref.~\cite{Putten-2023a} it has been identified a BH of $2.8 \ M_{\odot}$. If confirmed, the proposed mass gap will be narrowed to $2.4-2.8 \ M_{\odot}$,
assuming  $2.4 \ M_{\odot}$, as the upper limit for the mass of a NS. Furthermore, it is believed that the mass gap mentioned above could become more accurate with future observations of double neutron star (DNS) mergers, due to the variability in the mass distribution of DNS systems.~\cite{Putten-2023b}.

In any case, the  existence or not of the mass gap between NSs and BHs still remain an open problems. While some recent systematic studies support the existence of this gap, there are also recent observations of objects located within this area.  
The first event were the mass gap issue started to get interesting was the GW190814~\cite{Abbott-2020a} reporting an object with mass $2.6 \ M_{\odot}$, later another one, the GW190425~\cite{Abbott-2020-2} reports a compact binary coalescence with total mass $3.4 \ M_{\odot}$.
Very recently, the authors in Ref.~\cite{Abac-2024} (GW230529 event),  report the observation of a coalescing compact binary with component masses $2.5–4.5 \ M_{\odot}$ and  $1.2–2.0 \ M_{\odot}$ (all measurements quoted at the $90 \%$ credible level). The discovery of this system suggests an increase in the anticipated rate of NS–BH mergers accompanied by electromagnetic counterparts and adds further support to the existence of compact objects within the supposed lower mass gap.  Also, a number of recent studies (see also Ref.~\cite{Elise}) that have found evidence for compact objects within the mass gap, include observations of non-interacting binary systems~\cite{Todd-2018,Jayasinghe-2021}, radio pulsar surveys~\cite{Barr-2024}, and Gravitational Waves (GWs) from compact binary coalescence's~\cite{Abbott-2020a,Abbott-2023,Abbott-2017,Abbott-2020b}.

Regardless of the situation, an identification whether the component in the gap is a NS or a BH  would be remarkable. This distinction is crucial as it will help refine the boundaries of the mass gap, leading to positive advancements in our understanding of the physics of dense nuclear matter~\cite{Rezzolla-2018,Bailyn-1998,Farr-2011,Ozel-2010,Sa-2022,Tong-2021}. However, it has not yet been conclusively determined whether these components are BHs, NSs, or some other type of object. Future, more detailed observations may potentially shed light on which of these objects.

In particular, if this gap truly exists, a theoretical explanation for its presence must be provided. On the other hand, the potential existence of compact objects within this region must also be accounted for. Some compact objects observed in gravitational wave events have masses within the gap between known NSs and BHs. The nature of these mass-gap objects is still unknown, as is the formation process of their host binary systems (see also the related discussion in  Ref.~\cite{Elise}). In fact, there are some theoretical suggestions, including the possibility of exotic compact objects, that could explain these components, such as gravastars~\cite{Mazur-2004}, boson stars~\cite{Liebling-2023}, or Planck-scale modifications of BH horizons~\cite{Lunin-2002,Pani-2016}, which could also fall into this gap. 

Primordial BHs, which may have formed from over-dense regions in the early Universe, have also been suggested as potential candidates to populate the lower mass gap~\cite{Carr-2016}. In addition to astrophysical and primordial BHs within General Relativity (GR) and exotic compact objects, NSs predicted by alternative theories of GR could also occupy this mass gap. For example, in axionic scalar-tensor theory with viable phenomenological equations of state (EoS), such NSs could fall within this range~\cite{Oikonomou-2024}. 

In the present work, we propose an alternative explanation involving compact objects within the mass gap region. More specifically, we consider that there could be a stable compact object consisting of NS coexistence with DM within the region in question. This idea, brings to mind the analogy of the DM in Galaxies as an admixture with the baryonic matter, playing a vital role to the rotation curve. Another significant question is how DM could integrate with ordinary matter in a NS. A well-explored possibility is through capture, as detailed in Refs.~\cite{Goldman-89,Kouvaris-08,Bertone-08,Lavallaz-10,Kouvaris-10,Brito-15,Carmeno-17,Bramante-2013,Bell-2013}. Numerous other possibilities for how celestial bodies, such as NSs, can accumulate DM and interact with neutron matter—nucleons have been explored in previous studies~\cite{Thong-23, Sulagna-23, Maxim-19}.  

In particular in Ref.~\cite{Thong-23} the authors consider that NSs near the Galactic center may exist in a dense DM environment and could accumulate local DM clouds if it interacts efficiently with neutrons. This makes them valuable for studying DM annihilation. In models with light mediators, DM may annihilate into mediators that escape the star and subsequently decay into neutrinos outside of it. The research assesses constraints on heavy DM theories (TeV to PeV) involving long-lived mediators that decay into neutrinos, using high-energy neutrino observatory data and future projections These observations complement terrestrial searches and explore otherwise unreachable DM parameter spaces~\cite{Thong-23}. 

In Ref.~\cite{Sulagna-23}, the authors suggest that DM from the Galactic halo can accumulate in NSs and potentially convert them into BHs with masses under $2.5 \ M_{\odot}$, assuming the DM particles are heavy, stable, and interact with nucleons. They show that the lack of gravitational wave detections from mergers involving such low-mass BHs can be used to place constraints on the interactions between non-annihilating DM particles and nucleons.

The authors in Ref.~\cite{ Maxim-19} explore a model of composite DM involving stable particles with a charge of $-2$ that bind with primordial helium ($He$) nuclei through Coulomb forces to form $OHe$ atoms. They investigate how these dark atoms could be captured by regular matter, suggesting the potential existence of stable, superheavy nuclei enriched with $O$ and $O$ nuclearites, where heavy $O$ DM fermions are electromagnetically bound with ordinary nuclear matter. The study also examines how the accumulation of $OHe$ atoms in stars might influence stellar evolution, broadening the range of indirect methods to probe composite DM~\cite{ Maxim-19}.
Relevant discussion about the nature, the self-interaction of DM and dark objects are presented
also in Refs.~\cite{Kaplan-09,Vikiaris-2024,Boucenna-14,Zurek-14,Kouvaris-15,Arkani-09,Ray-2023,Maselli-2017,Nelson-19,Narain-06,Agnihotri-2009, Deliyergiyev-19,Zhang-2023,Freese-2016,Barbat-2024,Mariani-2024,Karvevandi-2022,Karvevandi-2024,Goldman-2013,SenD-2021,SenD-2024,SenD-2022,Sun-2024,AnnaWatts-2024,Husain-2023,Husain-2022}. 

In view of the above, a critical and logical question emerges: How can we distinguish these hybrid objects? What sets them apart from a classical NS or a pure DM star? Could the detection of associated gravitational waves (such as through tidal deformability) aid in their identification?
The formation and therefore the existence of these objects, could be justified and only astrophysical observations remains to confirm or refute their existence. In the present study, we also propose some possible ways that may lead to their observational identification.

The paper is organized as follows: In Section II, we introduce  the theoretical model
which encompasses the equations of state for both DM and NS matter, as well as the stability conditions for the hybrid objects. In Section III we present and discuss the results of this  study while in Section V, we conclude with our final remarks.

\section{The theoretical model}

\subsection{Two fluid model}
We consider a compact hybrid object made up of two fluids: non-self-annihilating DM particles mixed with NS matter. The composition of NSs is assumed to be in the hadronic phase, predominantly consisting of neutrons, with a small proportion of protons and electrons. Additionally, we assume that the two fluids, which together describe the total matter, interact exclusively via gravity. To predict the macroscopic properties of objects composed of these two fluids, it is necessary to solve the coupled Tolman-Oppenheimer-Volkoff (TOV) equations, with two equations for each fluid, simultaneously. These four equations are outlined below~\cite{Kodama-72,Sandin-09}       
\begin{eqnarray}
\frac{dP_{\rm NS}(r)}{dr}&=&-\frac{G{\cal E}_{\rm NS}(r) M(r)}{c^2r^2}\left(1+\frac{P_{\rm NS}(r)}{{\cal E}_{\rm NS}(r)}\right) \nonumber\\
&\times&
 \left(1+\frac{4\pi P(r) r^3}{M(r)c^2}\right) \left(1-\frac{2GM(r)}{c^2r}\right)^{-1}
\label{TOV-1}
\end{eqnarray}
\begin{equation}
\frac{dM_{\rm NS}(r)}{dr}=\frac{4\pi r^2}{c^2}{\cal E}_{\rm NS}(r)
\label{TOV-2}
\end{equation}
\begin{eqnarray}
\frac{dP_{\rm DM}(r)}{dr}&=&-\frac{G{\cal E}_{\rm DM}(r) M(r)}{c^2r^2}\left(1+\frac{P_{\rm DM}(r)}{{\cal E}_{\rm DM}(r)}\right) \nonumber \\
&\times&
 \left(1+\frac{4\pi P(r) r^3}{M(r)c^2}\right) \left(1-\frac{2GM(r)}{c^2r}\right)^{-1}
\label{TOV-1}
\end{eqnarray}
\begin{equation}
\frac{dM_{\rm DM}(r)}{dr}=\frac{4\pi r^2}{c^2}{\cal E}_{\rm DM}(r)
\label{TOV-2}
\end{equation}
where the subscripts $\rm NS$ and $\rm DM$ stand for the NS and DM respectively. Moreover  $P_i(r)$, ${\cal E}_i(r)$, $M_i(r)$ ($i={\rm NS}, {\rm DM}$) are the specific values of  pressure, energy density and mass.
Finally, the total pressure, energy density and mass are given by  $P(r)=P_{\rm NS}(r)+P_{\rm DM}(r)$,
 ${\cal E}(r)={\cal E}_{\rm NS}(r)+{\cal E}_{\rm DM}(r)$ and $M(r)=M_{\rm NS}(r)+M_{\rm DM}(r)$ respectively. To solve the TOV equations mentioned above, it is essential to know  the equations of state that characterize both NS matter (hadronic matter) and DM. The next two subsections focus on this topic.  

\subsection{Hadronic Equation of State}

In this study, to describe NS matter, we employ the equation of state  developed by Wang et al.~\cite{Wang-2022a} within the framework of Relativistic Brueckner-Hartree-Fock (RBHF) theory. This EoS has a microscopic foundation and its predictions align well with both the observed maximum masses and certain astrophysical constraints on radii~\cite{Wang-2022a,Laskos-2024}.

To be more specific, the NS Interior Composition Explorer (NICER) mission recently reported two independent Bayesian parameter estimates for the mass and equatorial radius of the millisecond pulsar PSR J0030+0451~\cite{NICER-1}. The predictions from the RBHF theory are fully consistent with the constraints provided by NICER~\cite{Wang-2022a}. Furthermore, the maximum NS masses $M_{\rm max}$ obtained from the RBHF theory align with the available astrophysical constraints from observations of massive NSs, such as PSR J1614-2230~\cite{Demorest-2016,Fonseca-2016},
PSR J0348+0432~\cite{Antoniadis-2013}, and PSR J0740+6620~\cite{Cromartie-2020,Fonseca-2021}.

\subsection{Dark matter model}
Regarding the DM particles, we assume they are relativistic fermions that interact with each other through a repulsive force. We consider a Yukawa-type interaction for this purpose~\cite{Nelson-19}
\begin{equation}
V(r)=\frac{{\rm g}_{\chi}^2 (\hbar c)}{4\pi r} \exp\left[-\frac{m_{\phi}c^2}{\hbar c}r\right]
\label{Yukawa-1}
\end{equation}
where ${\rm g}_{\chi}$  and $m_{\phi}$ are   the coupling constant  and  the mediator mass respectively. The interaction plays a crucial role in the formation and stability of mass gap objects. The contribution on the energy density of the self-interaction is given by
\begin{equation}
{\cal E}_{SI}(n_{\chi})=\frac{y^2}{2} (\hbar c)^3 n_{\chi}^2  
\end{equation}
where $y={\rm g}_{\chi}/m_{\phi}c^2$ (in units MeV$^{-1}$).
In this case the  total energy density of the DM particles is given by~\cite{Narain-06}
\begin{eqnarray}
{\cal E}_{\rm DM}(n_{\chi})&=&\frac{(m_{\chi}c^2)^4}{(\hbar c)^38\pi^2}\left[x\sqrt{1+x^2}(1+2x^2) \right. \nonumber\\
&-&\left. \ln(x+\sqrt{1+x^2})\right] + \frac{y^2}{2} (\hbar c)^3 n_{\chi}^2
\label{Rel-ED}
\end{eqnarray} 
The corresponding  pressure is calculated straightforward by using the definition
\begin{equation}
P_{\rm DM}(n_{\chi})=n_{\chi}\frac{d {\cal E}_{\rm DM}(n_{\chi})}{d n_{\chi}}-{\cal E}_{\rm DM}(n_{\chi})
\label{chi}
\end{equation}
and is given by
\begin{eqnarray}
P_{\rm DM}(n_{\chi})&=&\frac{(m_{\chi}c^2)^4}{(\hbar c)^38\pi^2}\left[x\sqrt{1+x^2}(2x^2/3-1)
\right. \nonumber \\
&+&\left. \ln(x+\sqrt{1+x^2})\right]  + \frac{y^2}{2} (\hbar c)^3 n_{\chi}^2
\label{Rel-Pr}
\end{eqnarray}
where $m_{\chi}$ is the particle mass and 
\[x=\frac{(\hbar c)(3\pi^2n_{\chi})^{1/3}}{m_{\chi}c^2}.  \]
The total energy density and pressure of the two-fluid mixing, considering that the contribution of NS matter are defined as ${\cal E}_{\rm NS}(n_b)$  and $P_{\rm NS}(n_b)$ where both are functions of the baryon density $n_b$, are given simply by the following sums 
\begin{equation}
{\cal E}(n_b,n_{\chi})={\cal E}_{\rm NS}(n_b)+{\cal E}_{\rm DM}(n_{\chi})
\label{eos-1-kouv}
\end{equation}
\begin{equation}
P(n_b,n_{\chi})=P_{\rm NS}(n_b)+P_{\rm DM}(n_{\chi}).
\label{P-tot-1}
\end{equation}

As we mentioned before, the DM is a self-interacting Fermi gas, where the contribution on the energy density of the self-interaction is given by $\frac{y^2}{2} (\hbar c)^3 n_{\chi}^2$, where $n_{\chi}$ is the DM density and $y={\rm g}_{\chi}/m_{\phi}c^2$ (in units MeV$^{-1}$), is the interaction strength. The interaction strength $y$ can be constrained by observational limits. These limits imposed on the cross section of the self-interaction~\cite{Burgio-2024,Liu-2024b}. In particular, according to \cite{Mark-2004,Kapli-2016,Sagun-2021,Loeb-2022} it holds $\sigma/ m_{\chi}\sim 0.1-10 \ {\rm cm}^2 /{\rm g}$. Moreover,  it has been showed that  the Born approximation is very accurate in the region $m_{\chi} \sim 1 \ {\rm GeV}~$\cite{Kouvaris-15,Maselli-2017,Tulin-2013}. Thus, we found that the interaction parameter $y$  varies in the range $\sim (0.001-0.1)({\rm GeV}/ m_{\chi}c^2)^{1/4} ({\rm MeV}^{-1})$~\cite{Burgio-2024}.

\subsection{Tidal deformability} 
 
A crucial source for gravitational wave detectors comes from the gravitational waves generated during the late phase of the inspiral of a binary NS system, just before the merger~\cite{Postnikov-2010,Flanagan-08,Hinderer-08}. This kind of source leads to the measurement of various properties of NSs. In the inspiral phase the tidal effects can be detected~\cite{Flanagan-08}.

The $k_2$ parameter, also known as tidal Love number, depends on the equation of state and describes the response of a NS to the tidal field $E_{ij}$~\cite{Flanagan-08}. The exact relation is given below
\begin{equation}
Q_{ij}=-\frac{2}{3}k_2\frac{R^5}{G}E_{ij}\equiv- \lambda E_{ij},
\label{Love-1}
\end{equation}
where $R$ is the radius of the compact object  and $\lambda=2R^5k_2/3G$ is the tidal deformability. The tidal Love number $k_2$ is given by \cite{Flanagan-08,Hinderer-08}
\begin{eqnarray}
k_2&=&\frac{8\beta^5}{5}\left(1-2\beta\right)^2\left[2-y_R+(y_R-1)2\beta \right]\nonumber\\
& \times&
\left[\frac{}{} 2\beta \left(6  -3y_R+3\beta (5y_R-8)\right) \right. \nonumber \\
&+& 4\beta^3 \left.  \left(13-11y_R+\beta(3y_R-2)+2\beta^2(1+y_R)\right)\frac{}{} \right.\nonumber \\
&+& \left. 3\left(1-2\beta \right)^2\left[2-y_R+2\beta(y_R-1)\right] {\rm ln}\left(1-2\beta\right)\right]^{-1},
\label{k2-def}
\end{eqnarray}
where $\beta=GM/Rc^2$ is the compactness of the star. The parameter $y_R$ is determined by solving numerically the following differential equation (for a detailed analysis for the case of the two fluid model see also Ref.~\cite{Hippert-2023}) 
\begin{equation}
r\frac{dy(r)}{dr}+y^2(r)+y(r)F(r)+r^2Q(r)=0. 
\label{D-y-1}
\end{equation}
$F(r)$ and $Q(r)$ are functionals of the total energy density ${\cal E}(r)$, total pressure $P(r)$, and total mass $M(r)$ defined as~\cite{Postnikov-2010,Hippert-2023}
\begin{equation}
F(r)=\left[ 1- \frac{4\pi r^2 G}{c^4}\left({\cal E} (r)-P(r) \right)\right]\left(1-\frac{2M(r)G}{rc^2}  \right)^{-1},
\label{Fr-1}
\end{equation}
and
\begin{eqnarray}
r^2Q(r)&=&\frac{4\pi r^2 G}{c^4} \left[\frac{}{} 5{\cal E} (r)+9P(r) \right. \nonumber\\
&+&\left. \frac{{\cal E}_{\rm NS} (r)+P_{\rm NS}(r)}{\partial P_{\rm NS}(r)/\partial{\cal E}_{\rm NS} (r)}+\frac{{\cal E}_{\rm DM} (r)+P_{\rm DM}(r)}{\partial P_{\rm DM}(r)/\partial{\cal E}_{\rm DM} (r)}\right]
\nonumber\\
&\times&
\left(1-\frac{2M(r)G}{rc^2}  \right)^{-1}- 6\left(1-\frac{2M(r)G}{rc^2}  \right)^{-1} \nonumber \\
&-&\frac{4M^2(r)G^2}{r^2c^4}\left(1+\frac{4\pi r^3 P(r)}{M(r)c^2}   \right)^2\left(1-\frac{2M(r)G}{rc^2}  \right)^{-2}.
\label{Qr-1}
\end{eqnarray}
Eq.~(\ref{D-y-1}) must be solved numerically and self-consistently with the coupled TOV equations, which correspond to the two-fluid model, under the following  conditions $y(0)=2$, $M_{NS}(0)=0$ and $M_{DM}(0)=0$  and with initial values of pressure the pair  $\{P_c^{\rm NS}, P_c^{\rm DM}$\}~\cite{Postnikov-2010,Hinderer-08,Hippert-2023}. From the numerical solution of TOV equations, the neutron and DM masses and radii can be computed, while the corresponding solution of the differential Eq.~(\ref{D-y-1}) provides the value of $y_R=y(R)$. The last parameter along with the quantity $\beta$ are the  basic ingredients  of the tidal Love number $k_2$.
Additionally, a related quantity in our study is   the dimensionless tidal deformability $\Lambda$  defined as  
\begin{equation}
\Lambda=\frac{2}{3}k_2 \left(\frac{c^2R}{GM}\right)^5.
\label{tidal-1}
\end{equation}
$\Lambda$  is a key parameter that can be determined through the detection of the associated gravitational waves and can provide valuable insights into both the structure and size of compact stars~\cite{Postnikov-2010,Hinderer-08,Hippert-2023}.

\subsection{Stability Conditions}

Ultimately, the essential research required involves examining the stability of these hybrid objects. Only after confirming their stability can we hypothesize that they may exist in nature. A common method for assessing stability is to analyze small radial perturbations of the equilibrium state by solving the Sturm-Liouville eigenvalue equation, which yields the eigenfrequencies~$\omega_n$~\cite{Shapiro-1983}. These eigenfrequencies form a discrete series where $\omega_n^2<  \omega_{n+1}^2$
$n=0,1,2\cdots$  where $\omega_n$ being real
numbers~\cite{Dengler-2022}. If $\omega_n^2$ is  negative, the radial perturbation grows exponentially, leading to the collapse of the star. The star is considered stable only when all eigenfrequencies are positive~\cite{Bardeen-1966} (see also the detailed discussion in Ref.~\cite{Shapiro-1983}). A notable  analysis of pulsation equations for compact stars consisting of multiple perfect fluids was recently conducted in Ref.~\cite{kain-2020,Hippert-2023}.  

In this study, we employ an alternative method to investigate the stability of these hybrid objects. Specifically, we use the method developed by Henriques, Liddle, and Moorhouse, which was applied to the study of boson-fermion stars~\cite{Henriques-1989,Henriques-1990a,Henriques-1990b}. This method has been developed and refined over the years for similar studies~\cite{Kain-21,Alvarado-2013,Alvarado-2020,Giovanni-2022,Hippert-2023}. Using this method, the stability analysis is performed by examining the behavior of baryons and dark DM particles, while keeping the total mass $M$ fixed. Specifically, the stability curve is constructed using the pair of central pressure values $\{P_c^{\rm NS}, P_c^{\rm DM}$\} corresponding to the points where the number of particles reaches its minimum and maximum values. These critical curves identify the transition between linear, stable, and unstable states with respect to perturbations that conserve both mass and particle number. Therefore, they satisfy the following conditions~\cite{Giovanni-2022}
\begin{eqnarray}
&&\left(\frac{\partial N_b}{\partial P_c^{\rm NS}}\right)_{\rm M=const}=\left(\frac{\partial N_{\chi}}{\partial P_c^{\rm NS}}\right)_{\rm M=const}=0 \nonumber \\
&&\left(\frac{\partial N_b}{\partial P_c^{\rm DM}}\right)_{\rm M=const}=\left(\frac{\partial N_{\chi}}{\partial P_c^{\rm DM}}\right)_{\rm M=const}=0
\end{eqnarray}
where $N_b$ and $N_{\chi}$ represent the numbers of baryons and DM particles, respectively. 
 These
numbers are given by the 
expression~\cite{Gresham-19}
\begin{equation}
N_{i}=4\pi\int_{0}^{R_i} n_i(r) \frac{r^2 dr}{\sqrt{1-2GM(r)/rc^2}}, \quad i=b,\chi
\end{equation}
where $R_i$ is the radius of the corresponding fluid.

The stability analysis can be summarized as follows~\cite{Giovanni-2022,Hippert-2023}: Sequences of stable equilibrium configurations starting from a purely DM star exhibit the characteristic that the number of DM particles  $N_{\chi}$ initially decreases
(as a function of the central pressures $P_c^{\rm DM}$ or $P_c^{\rm NS}$) until reaching the critical point, while the number of baryons $N_b$ increases. In general, equilibrium sequences may include continuous regions of both stability and instability. Therefore, in this case, multiple branches of stability can exist.

\section{Results and Discussion}

The number of configurations involving NS and DM mixing in the mass-gap region appears to be infinite. However, in this study, we are limited to only a few cases that can effectively validate our initial hypothesis. The two key parameters of the problem, which remain experimentally undetermined, are the potential mass of DM and the strength of its self-interaction. In fact, the range of these values spans several orders of magnitude for each parameter, and the type of DM - whether fermionic or bosonic - has yet to be determined. However, as we will demonstrate, the overlap range of the two parameters mentioned above, which result in mass-gap configurations, is quite extensive (we also worked with bosonic DM but due to the limitation in length, we present only the fermionic case). A more precise experimental determination of these parameters may also impose limits on the production of such configurations. 

\begin{figure}[h]
\centering  
\includegraphics[width=90mm, height=70mm]{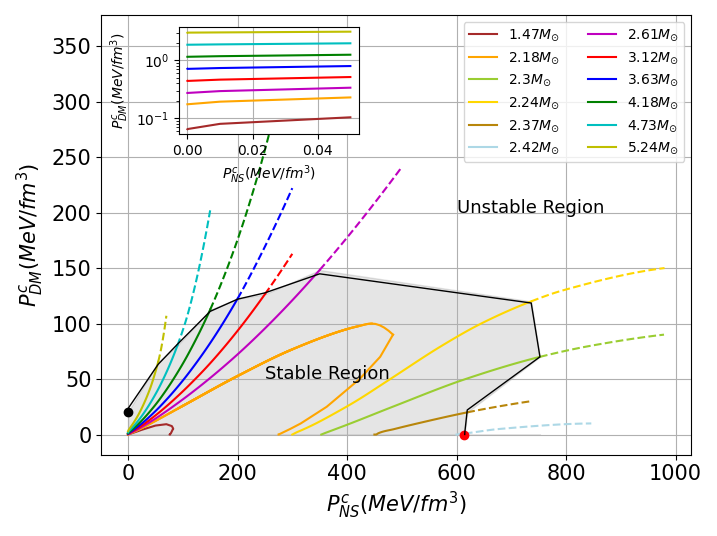}\\
\caption{Equilibrium configurations of NS and DM admixture, as a
function of the central pressures $P_{\rm DM}^c$ and $P_{\rm NS}^c$,  with interaction strength $y=0.05$ MeV$^{-1}$ and DM mass $m_{\chi}=1500$ MeV. The colored lines represent models with the same total mass, while the black solid line indicates the boundary between the stable and unstable regions. The inner figure visualizes more clearly the stable region for very low values of the central pressures.}
\label{equilb-1}
\end{figure}
\begin{figure}[h]
\centering  
\includegraphics[width=90mm, height=70mm]{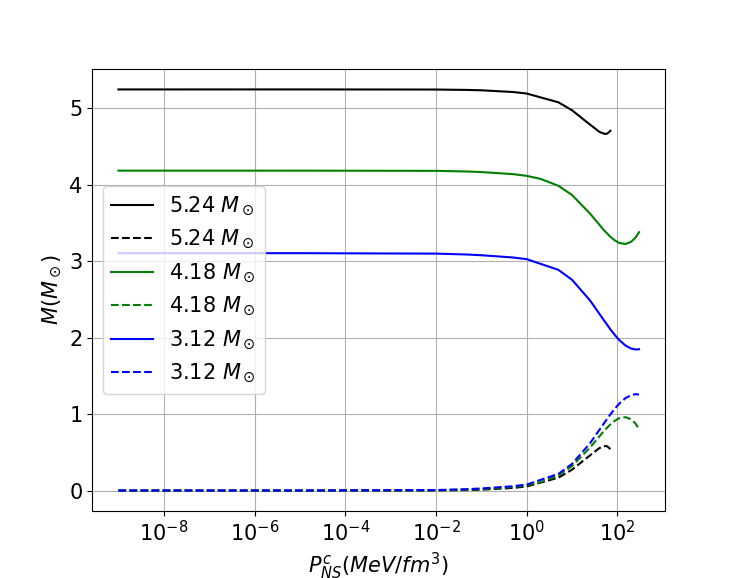}\\
\caption{The figure depicts the mass for the DM component (solid lines), separately with the NS component (dashed lines), of the compact object as a function of the central pressure of the NS matter $P_{\rm NS}^c$, for a three fixed values of the hybrid compact object.  }
\label{mass}
\end{figure}
\begin{figure}[h]
\centering  
\includegraphics[width=75mm, height=45mm]{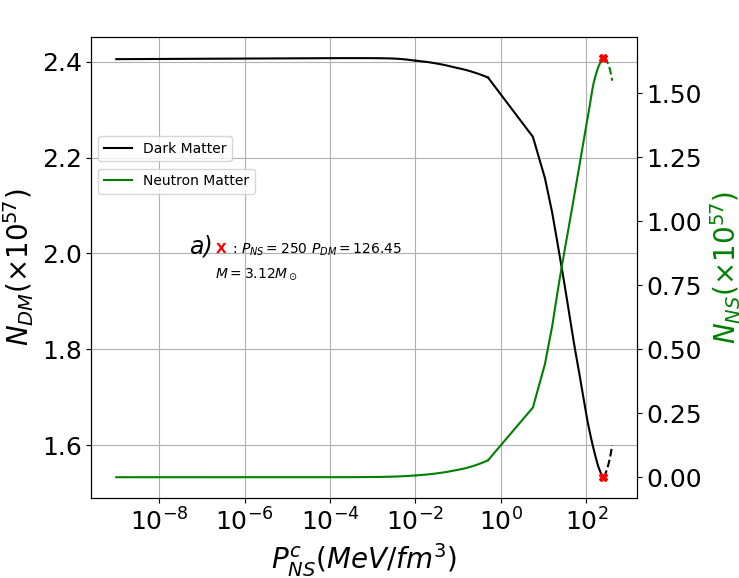}\\
\includegraphics[width=75mm, height=45mm]{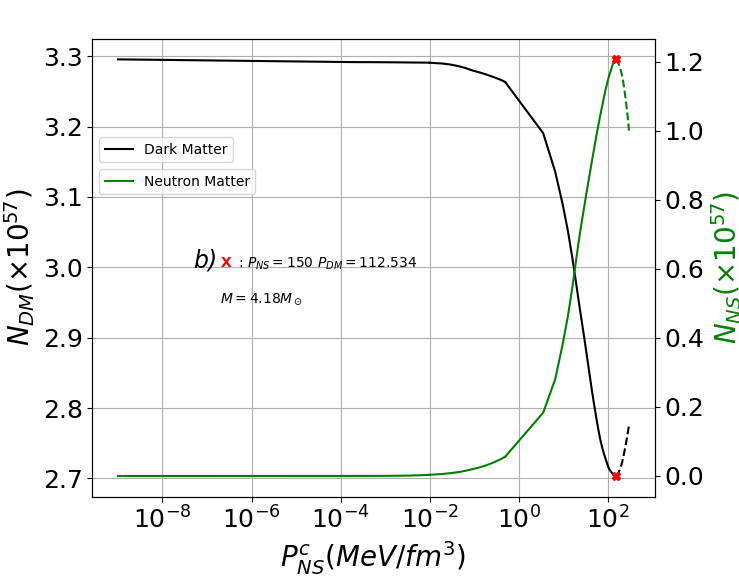}\\
\includegraphics[width=75mm, height=45mm]{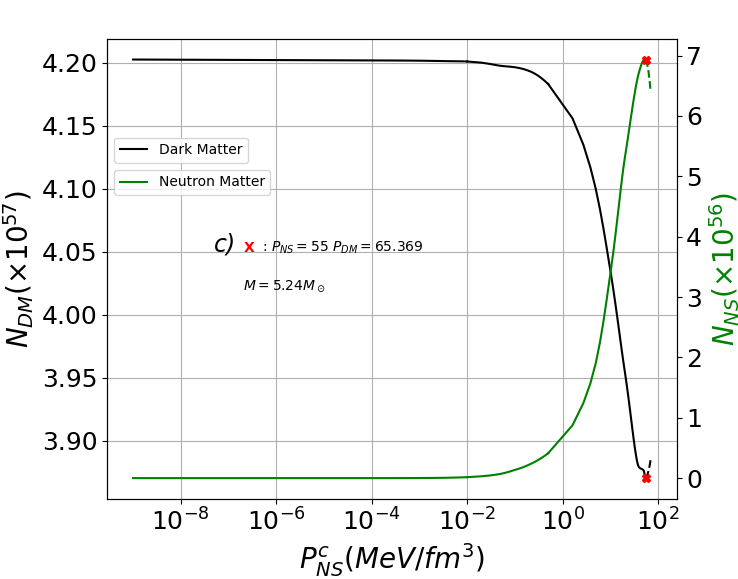}
\caption{The dependence of the number of DM particles $N_{\rm DM}$ and baryons $N_{\rm NS}$  on  the pressure    $P_{\rm NS}^c$  for equilibrium configurations of equal mass: a)  $M=3.12 \ M_{\odot}$, b)  $M=4.18 \ M_{\odot}$, c)  $M=5.24 \ M_{\odot}$. 
The solid lines indicate the stable branches, while the dashed lines are the unstable ones. The x appearing on each curve denotes the
last stable configuration (the corresponding central pressures, in ${\rm MeV/fm}^3$, are also indicated in each case).   }
\label{Number}
\end{figure}
\begin{figure}[h]
\centering  
\includegraphics[width=90mm, height=70mm]{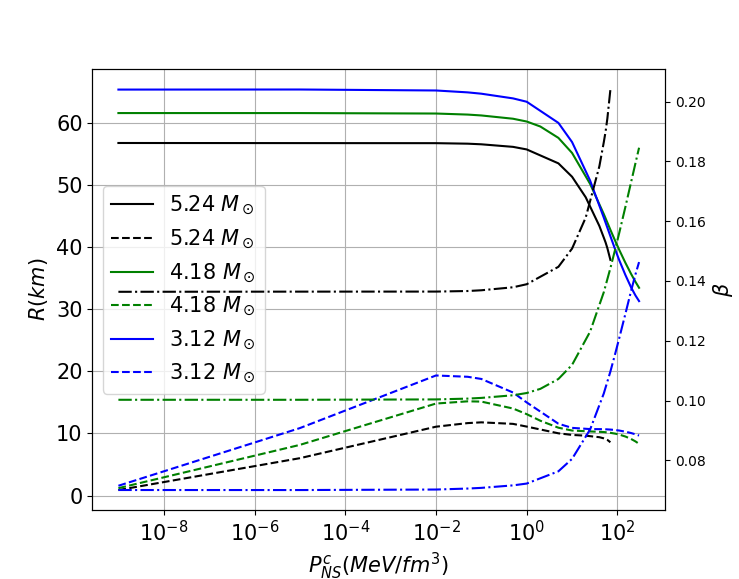}\\
\caption{The figure depicts the radius for the DM component (solid lines), separately with the NS component (dashed lines) and the compactness $\beta$ (dash-doted lines), of the compact object as a function of the central pressure of the NS matter $P_{\rm NS}^c$, for a three fixed values of the hybrid compact object.  }
\label{Radius}
\end{figure}
\begin{figure}[h]
\centering  
\includegraphics[width=90mm, height=70mm]{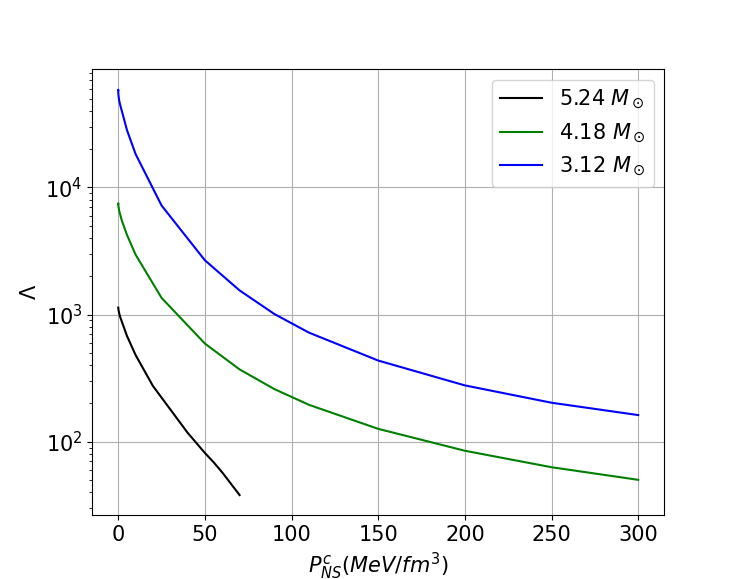}\\
\caption{The dimensional tidal deformability $\Lambda$ of the hybrid compact object as a function of the central pressure of the NS matter $P_{\rm NS}^c$, for a three fixed values of the hybrid compact object.}
\label{Tidal}
\end{figure}

To be more specific, in Fig.~\ref{equilb-1}, we present the equilibrium configurations of a NS with DM admixture and an interaction strength of $y=0.05$ MeV$^{-1}$ and a DM mass of $m_{\chi}=1500$ MeV. The different colored lines represent models with the same total mass, while the black solid line delineates the boundary between the stable and unstable regions. The black and red dots correspond to the cases of pure neutron star and pure DM star, respectively, along with their corresponding masses. This figure clearly demonstrates that the right combination of NS matter and DM can result in stable compact objects that could exist inside the mass-gap range. 
 
In Fig.~\ref{mass}, we illustrate the relationship between the individual mass of a NS and DM for three cases of hybrid objects, plotted as a function of central pressure of the NS matter $P_{\rm NS}^c$. These cases represent stable states corresponding to fixed total masses of $3.12$, $4.18$, and $5.24 \ M_{\odot}$. As expected, for small values of the hadronic pressure (and correspondingly high values of DM pressure) at the center of the hybrid object, the proportion of DM is overwhelmingly greater compared to that of the NS. However, as the hadronic pressure increases, the proportion of DM decreases, and the NS of the hybrid object approaches its known dimensions in terms of mass and radius. It is worth emphasizing that all these configurations describe stable states. Obviously, these results are sensitive to the fundamental parameters of the problem, which are the mass of the DM and the strength of its interaction. 

Another clear conclusion is that, within the aforementioned region, the larger the mass of the star, the higher the proportion of DM it contains. This is expected because, while the
NS equation of state limits the maximum mass of a NS, no such restriction applies to a DM star. Depending on the  mass of the DM particle and the strength of self-interaction, a DM star's mass can grow significantly larger choosing the above two parameters appropriately. In any case, the main conclusion remains unchanged: the admixture of DM with hadronic matter can lead to the formation of stable objects within the mass-gap region. 

In Fig.~\ref{Number}, we confirm the stability of the three selected cases (see Fig.~\ref{mass}), with the help of the $P_{\rm NS}^c-N_{\rm DM, NS}$ diagram. In particular, we display the dependence of the number of DM particles $N_{\rm DM}$ and baryons $N_{\rm NS}$  on  the pressure    $P_{\rm NS}^c$  for equilibrium configurations of equal mass: a)  $M=3.12 \ M_{\odot}$, b)  $M=4.18 \ M_{\odot}$ and  c)  $M=5.24 \ M_{\odot}$. 
The solid lines indicate the stable branches, while the dashed lines are the unstable ones. The x appearing on each curve denotes the
last stable configuration.

Given the importance of their dimensions, particularly for detecting these objects, Fig.~\ref{Radius} illustrates the relationship between the radius of these objects and the central pressure for relatively stable states with defined masses. Specifically, we plot the radius of the DM and NS components separately (with the total radius of the star corresponding to that of the DM component). The radius of the NS core falls within the typical range, while the radius of the DM component is approximately ten times larger. Nevertheless, despite their relatively large halos, these objects remain compact. Specifically, their compactness $\beta=GM/Rc^2$, as shown in Fig.~\ref{Radius}, is comparable to that of NSs.

However, since the radius of these stars is not easily detectable due to the formation of the DM halo, we focus on the tidal deformability of the star.
Thus, in Fig.~\ref{Tidal}, we plot the dimensional $\Lambda$ for three different masses.
Obviously, the values it takes are significantly larger compared to those of NSs. Since $\Lambda$ is a quantity that can be determined from observations, it serves as an important tool for detecting these objects. It is important to note that, for a given mass, the range of $\Lambda$ values can span 2 to 3 orders of magnitude. This outcome arises from the fact that these objects, in their stable state, can vary significantly in size, particularly in terms of their radius.

Moreover, we try to shed light on two critical issues concerning the creation and detection of these objects. {\it  How and when could supermassive compact objects composed of exotic fermions have formed?} A relevant discussion is provided in Ref.~\cite{Dengler-2022}. One possibility is the accretion of DM  into NSs  via the primordial formation of DM clumps surrounded by ordinary matter. Another possibility involves the formation of dark compact objects resulting from DM perturbations that grow from primordial over-densities~\cite{Chang-2019}. 

Finally, there are other possibilities for creating DM-admixed NSs with a significantly larger fraction of DM, which could accumulate gradually through capture over the NSs lifetime. One intriguing possibility is the substantial production and capture of DM during the core-collapse supernova of the neutron stars progenitor~\cite{Nelson-19,Gresham-19}.  In any case, there is no solid theoretical prediction that definitively rules out the formation of these hybrid NS-DM objects, where DM makes up a significantly larger proportion compared to NS matter.

The last proposition leads to the most critical question:  {\it how could we identify and detect these objects?} 
In this case, the most powerful tool is gravitational lensing, which takes advantage of the space-time distortion caused by these objects. Another possibility is detecting a potential merger event using the well-known detectors of the LIGO-Virgo-KAGRA Collaboration.
This includes two scenarios: a) the merger of two dark compact objects, or b) the merger of one dark compact object with another compact object, such as a NS  or a BH.
The corresponding GW signals will convey valuable information about the bulk properties of these objects-structures, including the average tidal deformability, individual masses, and chirp masses, offering insights into their structure and possibly some information about the distribution of DM in these objects~\cite{Bauswein-2023}.

Another potential detection scenario, as suggested in Ref.~\cite{Giovanni-2022}, involves a dark star in the second branch that has accreted a small amount of NS matter. Additionally, the star may have lost part of its bosonic matter due to accretion onto a second, more compact object. This process could push the star into the first unstable branch, ultimately activating the dispersion mechanism~\cite{Giovanni-2022}. 
Moreover, an extensive related discussion is given in Ref.~\cite{Hippert-2023}. In this study, the authors suggested  several ways to detect the presence of DM within NSs  and provided a list of the unique signatures associated with various scenarios.

\section{Concluding Remarks} 

Summarizing this study, we propose an alternative way to explain the possible existence of compact objects in the mass-gap region. In particular, we consider that there may be hybrid states of coexisting DM and hadronic matter in the form of stable compact stars in the universe, which, among other things, cover the mass-gap region.
To the best of our knowledge, this is the first attempt to explain the mass-gap using this particular hypothesis.
The formation of these objects allows for multiple interpretations, and despite its significance, it lies outside the scope of this study. 

Furthermore, we propose potential methods for observationally identifying these objects. While it is clear that more advanced detection techniques than those currently available are needed, we are confident that future advancements in this field will be capable of evaluating the validity of our hypothesis.

\section*{Acknowledgments}
This work was supported by the Czech Science Foundation (GACR Contract No. 21-24281S).


\end{document}